\documentclass[reprint,aps,nofootinbib,nobalancelastpage]{revtex4-2}
\usepackage{amsmath}
\usepackage{amsfonts}
\usepackage{amssymb}
\usepackage{amsthm}
\usepackage{mathtools}
\usepackage{graphicx}
\usepackage[dvipsnames]{xcolor}
\usepackage[colorlinks=True,citecolor=myBlue,linkcolor=myRed,urlcolor=myBlue,hypertexnames=false]{hyperref}
\usepackage{hypcap}
\usepackage{yfonts}
\usepackage{bm}
\usepackage[normalem]{ulem}
\usepackage{dsfont}
\usepackage{braket}
\usepackage{float}
\usepackage{marginnote}
\usepackage{bbding}
\usepackage{tikz}
\usepackage{quantikz}
\usetikzlibrary{shapes, arrows.meta,decorations.pathreplacing}

\DeclarePairedDelimiter{\norm}{\lVert}{\rVert} 
\usepackage{kantlipsum}

\newcommand\eq[1]{\begin{align}#1\end{align}}

\newcommand\mytitle{Ensemble phase transitions in measurement-unitary hybrid quantum circuits}

\definecolor{myBlue}{RGB}{31,119,180}
\definecolor{myOrange}{RGB}{255,127,14}
\definecolor{myGreen}{RGB}{44,160,44}
\definecolor{myRed}{RGB}{214,39,40}
\definecolor{myPurple}{RGB}{148,103,189}

\begin{document}

\title{\mytitle}

\author{Saptarshi Mandal}
\email{saptarshi.mandal@icts.res.in}
\affiliation{International Centre for Theoretical Sciences, Tata Institute of Fundamental Research, Bengaluru 560089, India}

\author{Sthitadhi Roy}
\email{sthitadhi.roy@icts.res.in}
\affiliation{International Centre for Theoretical Sciences, Tata Institute of Fundamental Research, Bengaluru 560089, India}

\begin{abstract}
The universal properties of state ensembles generated by quantum dynamics provide a refined lens into Hilbert-space ergodicity and deep thermalisation, or lack thereof. We investigate deep-ergodicity breaking transitions of an ancillary qubit entangled with a system undergoing dynamics via hybrid quantum circuits, where unitary evolution is interspersed with projective measurements. Specifically, we analyse the statistical properties of two distinct state ensembles: the trajectory ensemble (TE) conditioned on the measurement outcomes of the hybrid circuit, and the projected ensemble (PE) generated by further projective measurements of the system at the final temporal boundary. We find that both ensembles undergo non-analytic changes -- ensemble phase transitions -- at the critical point of the measurement-induced purification transition, despite their first moments remaining completely maximally mixed and blind to the transition. For Haar-random circuits, the TE transitions from a trivial identity ensemble to the uniform Haar ensemble, while the PE transitions from the Haar ensemble to a trivial singleton pure state. In Clifford circuits, absence of non-stabiliserness limits the PE to a $1$-design, while the TE forms a $k$-design for $k \leq 3$. We demonstrate that these genuine ensemble transitions offer a clear physical mechanism for the measurement-induced entanglement or purification transitions, illustrating how the capacity to ergodically steer a quantum subsystem shifts from the temporal boundary measurements in the mixed phase to the bulk trajectory measurements in the pure phase.
\end{abstract}

\maketitle

\section{Introduction \label{sec:intro}}

The universal properties of ensembles of states generated by quantum dynamics are central to addressing the fundamental questions of thermalisation and information scrambling in quantum statistical mechanics~\cite{popescu2006entanglement,rigol2008thermalisation,nandkishore2015many,dalessio2016from,hayden2007black}. Such ensembles are also key towards establishing the connections between questions in quantum many-body dynamics to those in quantum information theory such as emergence of quantum $t$-designs and quantum error correction~\cite{gross2007evenly,roberts2017chaos,cotler2023emergent,ho2022exact,choi2020quantum,Gullans2020Purification}.
Defining the ensemble of states as 
\eq{
{\cal E} = \{p(\varrho),~\varrho\}\,,
}
where $p(\varrho)$ is the probability of the state $\varrho$, it is its first moment $\rho \equiv \sum_{\varrho \in {\cal E}} p(\varrho)\varrho$ which is conventionally studied as it encodes, in principle, all the average correlations characterising the system. 
However, the higher moments, defined via
\eq{
\rho^{(k)}\equiv \sum_{\varrho\in {\cal E}}p(\varrho)\,\varrho^{\otimes k}\,,
}
manifestly contain fundamentally more information not present in the first moment. 
It is therefore natural to ask if there exist universal forms of the entire ensemble and how those forms are affected by the nature of the quantum dynamics generating them; in other words do there exist phase transitions of the ensembles as certain parameters characterising their generating dynamics is varied.

A setting where ensembles of states appear quite naturally are measurement-unitary hybrid dynamics where the unitary quantum dynamics are interspersed with measurements.
The probabilistic nature of the measurement outcomes generates an ensemble of states conditioned on the set of outcomes with the corresponding probabilities given by the Born-rule probabilities associated to the measurements. 
It is worth mentioning that while such models are of obvious theoretical interest as a rich playground for studying the interplay of entangling unitaries and disentangling measurements in the resulting ensembles, they are routinely realisable in the state-of-the-art quantum computing platforms including quantum gas microscopes~\cite{bakr2009gasmicroscope,sherson2010single}, trapped Rydberg-atom arrays~\cite{bernien2017simulator,ebadi2021quantum,norcia2023midcircuit,bluvstein2024logical}, trapped ions~\cite{noel2022measurement,moses2023trappedions,agrawal2024observing}, and superconducting qubit arrays~\cite{koh2023experimental,hoke2023measurement,kamakari2025experimental}. 

These measurements can be present mid-circuit during the dynamics in which case they generate ensembles of trajectories conditioned on the sequence of the outcomes or could be present at the end of the unitary dynamics where the outcomes generate an ensemble of states on the subsystem complementary to the measured one. 
The former is central to the idea of measurement-induced entanglement transitions~\cite{li2018quantum,skinner2019measurement,li2019measurement,choi2020quantum,Gullans2020Purification,bao2020theory,jian2020measurement,zabalo2020critical,gullans2020scalable,nahum2021measurement,potter2022entanglement,fisher2023random,ippoliti2021entanglement,nahum2020entanglement,lavasani2021measurement,sang2021measurement,ippoliti2021postselection,cao2019entanglement,alberton2021entanglement,buchhold2021effective,poboiko2023theory,fava2023nonlinear,agarwal2022charge,barratt2022field,barrat2022charge} where the bipartite entanglement in individual trajectories of the dynamics (averaged over the trajectories) undergoes a transition from volume-law to area-law as the rate of measurements is increased. Importantly, the average (over the ensemble) density matrix of the system or any of its subsystems is blind to this transition. This reflects that the transition is truly a property of the ensemble encoded in its higher moments, which can only be accessed via quantities that are non-linear in the states in the ensemble.

On the other hand, ensembles of states generated on subsystems conditioned on projective measurement outcomes on the complementary subsystem, referred to as the {\it projected ensemble} (PE) have recently attracted a lot of attention. This is, in part, due to the fact that the PEs approach universal forms when generated from a single global state via maximum-entropy principles under the umbrella of Hilbert-space ergodicity and deep thermalisation~\cite{cotler2023emergent,choi2023preparing,ho2022exact,ippoliti2022solvablemodelofdeep,ippolitu2023dynamical,mark2024maximum,mcginley2025scroogeensemblemanybodyquantum,mok2026naturestingyuniversalityscrooge,mark2026deep}. 
This has fundamental implications for the emergence of designs or lack thereof, and provides a finer probe of ergodicity and the quantum information content of states beyond conventional bipartite entanglement measures~\cite{ippolitu2023dynamical,lucas2023freefermiondeepthm,bhore2023deepthmconstrained,varikuti2024unravelingemergence,chan2024pe,manna2025peconserved,chang2025charge,mark2023benchmarking,shaw2024universal,loio2025quantum}.

An overarching question in this context is how to understand the deep ergodicity of such ensembles, or lack thereof, from the point of view of their quantum resource content~\cite{chitambar2019quantum}. Recent works have shown that the PEs can show deep-ergodicity breaking transitions reflected in transitions of the resource content of the states in settings where the transition is driven by the amount of global resource injected initially and the dynamics preserves it~\cite{varikuti2025deep,liu2026coherence,feng2026quantum}.

However, hybrid circuits offer an interesting and unique situation where the measurements which generate the ensembles also control the amount of resource, namely entanglement, in the states. 
Motivated by this, we ask the question that what is the fate of deep thermalisation in the ensembles of trajectories, as well as the PE, as the rate of measurements is varied, and in particular, do these ensembles show non-analytic changes in their nature -- {\it ensemble phase transitions} -- and if they are related to the measurement-induced entanglement transitions in hybrid circuits.

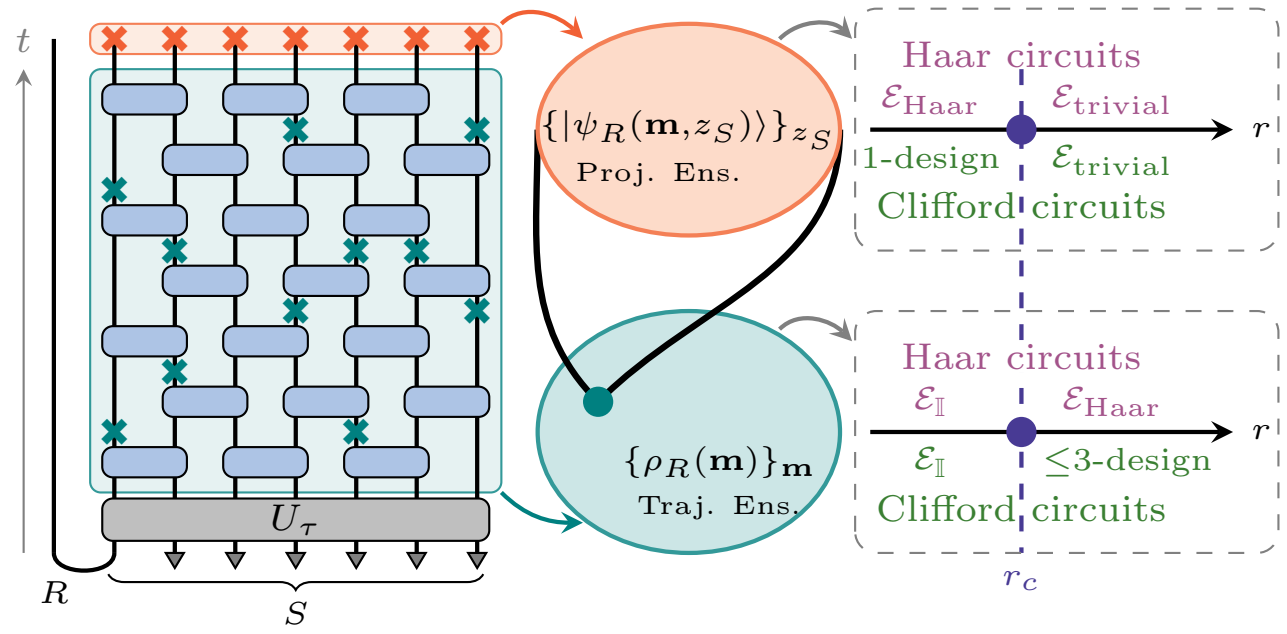
\begin{figure}[t]
\centering
\begin{tikzpicture}[scale=0.4]
% \foreach \x in {0,...,20}{
%     \draw (\x,0) -- (\x,10);
% }
% \foreach \x in {0,...,10}{
%     \draw (0,\x) -- (20,\x);
% }
% \draw[thick](8,0) -- (8,10);
% \draw[thick](13,0) -- (13,10);

\filldraw[fill=teal!10,draw=teal!80,rounded corners=2] (0.6,1.5) rectangle (7.4,8.5);
\draw[teal, thick, >=stealth, ->] (7.4,1.5) to[out=-45,in=180] (8.75,1);
\filldraw[fill=RedOrange!10,draw=RedOrange!80,rounded corners=2] (0.6,8.75) rectangle (7.4,9.25);
\draw[RedOrange, thick, >=stealth, ->] (7.4,9.25) to[out=45,in=135] (8.75,9.);

\foreach \x in {0,...,7}{
    \draw[thick] (\x,0.5) -- (\x,9);    
}
\foreach \x in {2,...,7}{
\filldraw[fill=gray,draw=black] (\x-0.15,0.5) -- (\x+0.15,0.5) -- (\x,0.25) --cycle;
\node at (\x,9) {\tiny{{\color{RedOrange}\XSolidBold}}};
}
\node at (1,9) {\tiny{{\color{RedOrange}\XSolidBold}}};
\draw[thick] (0,0.5) to[out=270,in=270] (1,0.5);
\filldraw[fill=lightgray, draw=black, rounded corners=2] (0.8,0.7) rectangle (7.2,1.4);
\node at (4,1.05) {\scriptsize{$U_\tau$}};

\foreach \y in {2,4,6,8}{
\foreach \x in {1,3,5}{
\filldraw[fill=RoyalBlue!30, draw=black, rounded corners=2] (\x-0.2,\y-0.25) rectangle (\x+1.2,\y+0.25);
}
}
\foreach \y in {3,5,7}{
\foreach \x in {2,4,6}{
\filldraw[fill=RoyalBlue!30, draw=black, rounded corners=2] (\x-0.2,\y-0.25) rectangle (\x+1.2,\y+0.25);
}
}

\node at (1,2.5) {\tiny{{\color{teal}\XSolidBold}}};
\node at (5,2.5) {\tiny{{\color{teal}\XSolidBold}}};
\node at (2,3.5) {\tiny{{\color{teal}\XSolidBold}}};
\node at (4,4.5) {\tiny{{\color{teal}\XSolidBold}}};
\node at (7,4.5) {\tiny{{\color{teal}\XSolidBold}}};
\node at (2,5.5) {\tiny{{\color{teal}\XSolidBold}}};
\node at (5,5.5) {\tiny{{\color{teal}\XSolidBold}}};
\node at (6,5.5) {\tiny{{\color{teal}\XSolidBold}}};
\node at (1,6.5) {\tiny{{\color{teal}\XSolidBold}}};
\node at (4,7.5) {\tiny{{\color{teal}\XSolidBold}}};
\node at (7,7.5) {\tiny{{\color{teal}\XSolidBold}}};

% \foreach \x in {0,...,10}{
%     \draw (0,\x) -- (20,\x);
% }
% \draw[thick](8,0) -- (8,10);
% \draw[thick](13,0) -- (13,10);

\filldraw[fill=teal!20, draw=teal!80, thick] (10.5,2.5) ellipse (2.5 and 2);
\draw[very thick] (9,3) to[out=45, in=270] (13,7.5);
\draw[very thick] (9,3) to[out=135, in=270] (8,7.5);
\fill[teal] (9,3) circle (0.25);
\filldraw[fill=RedOrange!20, draw=RedOrange!80, thick] (10.5,7.5) ellipse (2.5 and 1.75);
\node at (11,2) {$\scriptstyle \{\rho_R({\bf m})\}_{\bf m}$};
\node at (11,1.25) {{\tiny{Traj. Ens.}}};
\node at (10.5,7.5) {$\scriptstyle {\{\ket{\psi_R({\bf m},z_S)}\}_{z_S}}$};
\node at (10,6.75) {{\tiny{Proj. Ens.}}};

\draw[thick, >=stealth, ->] (13.5,2.5) -- (19.5,2.5) node[right] {$\scriptstyle r$};
\node at (14.5,3) {\color{DarkOrchid}{$\scriptstyle{\cal E}_{\mathbb{I}}$}};
\node at (14.5,2) {\color{OliveGreen}{$\scriptstyle{\cal E}_{\mathbb{I}}$}};
\node at (17.5,3) {\color{DarkOrchid}{$\scriptstyle{\cal E}_{\mathrm{Haar}}$}};
\node at (17.75,2) {\color{OliveGreen}{$\scriptstyle \leq 3{\text{-design}}$}};
\draw[thick, >=stealth, ->] (13.5,7.5) -- (19.5,7.5) node[right] {$\scriptstyle r$};

\node at (14.5,8) {\color{DarkOrchid}{$\scriptstyle{\cal E}_{\mathrm{Haar}}$}};
\node at (14.5,7) {\color{OliveGreen}{$\scriptstyle 1{\text{-design}}$}};
\node at (17.5,8) {\color{DarkOrchid}{$\scriptstyle{\cal E}_{\mathrm{trivial}}$}};
\node at (17.5,7) {\color{OliveGreen}{$\scriptstyle {\cal E}_{\rm trivial}$}};

\draw[BlueViolet, thick,dashed] (16,8.5) -- (16,0.5) node[below] {$\scriptstyle r_c$};

\fill[BlueViolet] (16,7.5) circle (0.25); 
\fill[BlueViolet] (16,2.5) circle (0.25);

\node at (0,-0.15) {$\scriptstyle R$};
\draw[decorate, decoration={brace, mirror, amplitude=3pt}] (0.9,0.15) -- (7.1,0.15)
    node[midway, below=2pt] {$\scriptstyle S$};

\filldraw[fill=white,draw=white](15.5,3.5) rectangle (16.5,4.);
\node at (16,8.75) {\scriptsize{{\color{DarkOrchid}Haar circuits}}};
\node at (16,6.25) {\scriptsize{{\color{OliveGreen}Clifford circuits}}};

\node at (16,3.75) {\scriptsize{{\color{DarkOrchid}Haar circuits}}};
\node at (16,1.25) {\scriptsize{{\color{OliveGreen}Clifford circuits}}};
\draw[>=stealth,gray,->] (-0.5,0.5) -- ++(0,8) node[above] {\scriptsize{$t$}};

\draw[gray, dashed, rounded corners] (13.25,5.5) rectangle (20.25,9.5);
\draw[gray, dashed, rounded corners] (13.25,0.5) rectangle (20.25,4.5);

\draw[black!50, >=stealth, ->, thick] (12,9) to[out=45, in=135] (13.2,9);

\draw[black!50, >=stealth, ->, thick] (12,4.2) to[out=45, in=135] (13.2,4);

\end{tikzpicture}
\caption{Setting and summary of the main result of this work.
A reference qubit $R$ is prepared in a Bell pair 
with one of the $N$ qubits of the system $S$, following which $S$ is evolved with a scrambling 
warm-up circuit $U_\tau$ of depth $\tau=O(N)$. Thereafter, $S$ is evolved with a 
hybrid brickwork circuit of depth $t$, where the two-qubit unitary gates (blue) are 
interspersed with projective single-qubit measurements (green crosses) with rate $r$. The outcomes of these measurements generate trajectory ensemble. Each member of the trajectory ensemble is further unravelled into a projected ensemble of pure states in $R$ obtained by projectively measuring each qubit in $S$ (orange crosses). The trajectory and projected ensembles, both show ensemble phase transitions as a function of $r$ between ensembles mentioned in the figure. The critical point $r_c$ coincides with that of the MIPT.}
\label{fig:setting}
\end{figure}

The main result of our analyses is summarised in Fig.~\ref{fig:setting}.
We consider a setting where a reference qubit $R$ is entangled with system $S$ and the latter goes undergoes dynamics effected by a 1+1D brickwork hybrid circuit with a measurement $r$. 
This is known to host a measurement-induced purification transition (MIPT)~\cite{Gullans2020Purification,gullans2020scalable} for $R$ at a critical rate $r_c$.
Although $R$ is a single qubit, it acts as a local, scalable probe of the dynamics of $S$~\cite{gullans2020scalable}, and the picture below is expected to hold equally for a reference comprising a finite number of qubits.
The set of measurement outcomes, ${\bf m}$, in the circuit defines an ensemble of trajectories for each of which there exists a state $\rho_R({\bf m})$ obtained after tracing out $S$.
For each such trajectory, instead of tracing out $S$, all the qubits therein can be further measured, the set of outcomes for which generates a PE of pure states for $R$.
In this way, the two ensembles constitute finer probes of the dynamics --  the {\it trajectory ensemble} (TE) unravels the averaged dynamics into an ensemble of states conditioned on the bulk outcomes whereas the PE further unravels each member of the TE into an ensemble of pure states conditioned on the final measurement outcomes of $S$.

We find that the TE undergoes a phase transition at $r_c$. 
When the unitary gates are Haar-random, it transitions from an identity ensemble, wherein the ensemble consists solely of the maximally mixed state on $R$, to the Haar-ensemble where it forms state $k$-designs for arbitrary $k$. By contrast, if the gates are random Clifford, the ensemble transitions from the identity ensemble to a $k$-design for $k\leq 3$. 
The PE on the other hand, for Haar-random circuits, undergoes a transition from the Haar ensemble to a trivial ensemble wherein the entire PE consists of a single pure state. For Clifford circuits, given a trajectory, the states in the PE are constrained to lie along a fixed stabiliser axis with the final outcomes merely shuffling the sign along this direction. Consequently, the PE generates only a 1-design instead of the Haar ensemble before transitioning to the trivial ensemble.

It is important to note that the first moment of all the ensembles for all values of $r$ is given by the maximally mixed state and is completely blind to the transitions. 
The transitions are therefore genuinely at the level of the ensembles and constitute examples of deep-ergodicity breaking transitions beyond conventional ergodicity breaking transitions.
Crucially, the standard MIPT -- viewed here as a purification transition of the reference qubit $R$ -- can be naturally understood as the physical mechanism linking both ensemble transitions. In the mixed phase ($r < r_c$), the bulk measurements fail to purify $R$, leaving every trajectory state $\rho_R({\bf m})$ maximally mixed and thereby yielding the identity TE. However, because $R$ remains entangled with $S$, the final projective measurements on $S$ can ergodically steer $R$ over its entire Hilbert space, generating the Haar PE. Conversely, in the pure phase ($r > r_c$), the bulk measurements extract sufficient information to fully purify $R$, with the measurement outcomes steering $R$ across the Hilbert space to form the Haar TE. Consequently, because $R$ is already pure and disentangled from $S$ in each trajectory, the final measurements on $S$ impart no further backaction on $R$, forcing the PE to collapse into the trivial singleton ensemble.

The remainder of the paper is organised as follows. In Sec.~\ref{sec:models} we describe the model concretely and define in detail the two ensembles, the trajectory ensemble and the projected ensemble.
The transitions in the trajectory ensemble are presented in Sec.~\ref{sec:te}, whereas those for the PE are discussed in Sec.~\ref{sec:pe}.
In Sec.~\ref{sec:mipt} we discuss how the MIPT is a manifestation of the underlying ensemble transitions. We close with a summary and outlook in Sec.~\ref{sec:summary}. Additional details pertaining to calculations for the Clifford circuits and additional results for all-to-all hybrid circuits are relegated to Appendices~\ref{sec:sm-te} and \ref{sec:sm-alltoall}, respectively.

\section{Model and Ensembles of States \label{sec:models}}
As a concrete model, 
we consider the setting shown in Fig.~\ref{fig:setting}, 
where a reference qubit $R$ is initially entangled with an $N$-qubit system, $S$, with the latter undergoing hybrid dynamics.
More specifically,
$R$ is prepared in a Bell pair state with one of the $N$ qubits of $S$, and its information is scrambled across $S$ by a warm-up circuit $U_\tau$ of depth $\tau=O(N)$.
$S$ is then evolved with a hybrid brickwork circuit of depth $t$, wherein every qubit is projectively measured with probability $r$ after each layer of gates.
The specific gates and the locations of measurements define a circuit realisation ${\cal C}$.
In the rest of what follows, the ensembles will be defined for a given ${\cal C}$ and the average over circuit realisations will be denoted by $\mathbb{E}_{\cal C}[\cdot]$.

For a given ${\cal C}$ and
a sequence of 
measurement outcomes ${\bf m}$, the state of $R\cup S$ is
\eq{
\ket{\Psi_{RS}({\bf m})} = \frac{K_{\bf m}U_\tau\ket{\Phi^+}_{RS}}{\sqrt{p({\bf m})}}\,,
\label{eq:psi-traj}
}
where 
\eq{
\ket{\Phi^+}_{RS} = \frac{1}{\sqrt{2}}\left(\ket{0_R0_1}+\ket{1_R1_1}\right)\bigotimes_{i=2}^N\ket{0_i}
}

is the initial state with $R$ in a Bell-pair state with the qubit ${i=1}$ of $S$ and the remaining qubits of $S$ in a product state, $K_{\bf m}$ is the non-unitary operator implemented by the hybrid circuit and $p({\bf m})=\Vert K_{\bf m}U_\tau\ket{\Phi^+}_{RS}\Vert^2$ is the Born rule probability of 
obtaining the outcome sequence ${\bf m}$.

We now explicitly define the measurement-induced ensembles that we study in the rest of this work.
The sequence of measurement outcomes ${\bf m}$ characterising a trajectory of the hybrid dynamics naturally generates an ensemble of states over ${\bf m}$, generically mixed, on $R$ which we refer to as the {\it trajectory ensemble} (TE).
Formally, it is defined as
\eq{
{\cal E}_{\rm TE}\equiv\{p({\bf m}),\rho_R({\bf m})\}_{\bf m}
\,,
\label{eq:te-def}
}
where $\rho_R({\bf m})={\rm Tr}_S\ket{\Psi_{RS}({\bf m})}\bra{\Psi_{RS}({\bf m})}$.

Since each state in the TE is obtained by tracing out $S$, the TE does not contain any information about $S$ or 
about $R$ that is still stored in $S$.

However, each state $\rho_R({\bf m})$ can be further unravelled by measuring each qubit in $S$, which generates a yet finer ensemble of pure states on $R$.
This is precisely the setting of the {\it projected ensemble} (PE) and of deep thermalisation~\cite{cotler2023emergent,choi2023preparing}, and amounts, in experiments, to complementing the mid-circuit measurements with a final site-resolved snapshot of $S$.
Measuring $S$ at time $t$ in the computational basis thus resolves each member of the TE further, and leads, for a given ${\bf m}$, to a PE on $R$ conditioned on both the bulk outcomes ${\bf m}$ and the boundary outcomes $z_S$,
\eq{
{\cal E}_{\rm PE}({\bf m})\equiv\{p(z_S|{\bf m}),\ket{\psi_R({\bf m},z_S)}\}_{z_S}\,,
\label{eq:pe-def}
}
where the states in the PE are given by
\eq{
\ket{\psi_R({\bf m},z_S)}\otimes\ket{z_S}=\Pi_{z_S}\ket{\Psi_{RS}({\bf m})}/\sqrt{p(z_S|{\bf m})}
}
with 
$p(z_S|{\bf m})=\braket{\Psi_{RS}({\bf m})|\Pi_{z_S}|\Psi_{RS}({\bf m})}$ and 
$\Pi_{z_S}=\ket{z_S}\bra{z_S}$.
Before delving deeper into how the rate of measurements drives a transition in the nature of the ensembles, we make a few general remarks about the limiting behaviours in low and high measurement rates and define appropriate diagnostics for the transitions.

The average purity of the states in ${\cal E}_{\rm TE}$ is precisely the order parameter for the measurement-induced purification transition~\cite{Gullans2020Purification,gullans2020scalable}.
For low measurement rates, each state in the TE in Eq.~\eqref{eq:te-def} is maximally mixed such that $\rho_R({\bf m})=\mathbb{I}/2~\forall {\bf m}$; we refer to this as the {\it identity ensemble} ${\cal E}_{\mathbb{I}}$.
On the other hand, for high measurement rates, the hybrid dynamics purifies the state of $R$; however the specific pure state $\rho_R({\bf m})$ approaches can depend on ${\bf m}$ and therefore a question arises that what is the pure-state ensemble to which ${\cal E}_{\rm TE}$ approaches.
Since the dynamics has no conservation laws, it is expected that the first moment $\sum_{\bf m}p({\bf m})\rho_R({\bf m}) = \mathbb{I}/2$.
As such, the corresponding maximum-entropy pure-state ensemble is just the Haar ensemble.
This motivates us to study, as a function of $r$, the distance of the $k^{\rm th}$ moment of ${\cal E}_{\rm TE}$ from that of the Haar and the identity ensembles, quantified by the trace distance
\eq{
{\cal D}^{(k)}_{\cal X} = \frac{1}{2}\norm{\rho^{(k)}_{\rm TE}-\rho^{(k)}_{\cal X}}_1\,;~~{\cal X}\in\{{\rm Haar},\mathbb{I}\}\,,
\label{eq:Dk-te}
}
where $\norm{.}_1$ denotes the trace norm. Here $\rho^{(k)}_{\mathbb{I}}=(\mathbb{I}/2)^{\otimes k}$ and $\rho^{(k)}_{\rm Haar}=\Pi^{(k)}_{\rm sym}/(k+1)$, with $\Pi^{(k)}_{\rm sym}$ the projector onto the symmetric subspace of $k$ copies of $R$.
Since the first moments of all three ensembles coincide, these distances are non-trivial only for $k\geq 2$.

Turning to the PE, note that from the definition in Eq.~\eqref{eq:pe-def}, it is defined for each measurement trajectory ${\bf m}$, and hence, so is its moment
\eq{
\rho^{(k)}_{\rm PE}({\bf m})\!=\! \sum_{z_S}p(z_S|{\bf m})\left[\ket{\psi_R({\bf m},z_S)}\bra{\psi_R({\bf m},z_S)}\right]^{\otimes k}\!.
\label{eq:pe-moments}
}
In the limit of $r\to 0^+$, each member of the TE which is nothing but the first moment of the corresponding PE is proportional to identity; $\rho_R({\bf m})=\rho_{\rm PE}^{(k=1)}({\bf m})=\mathbb{I}/2$.
The PE unravels this maximally mixed state into an ensemble of, potentially random, pure states where the randomness arises from that of the measurement outcomes $z_S$. 
Since the first moment is $\mathbb{I}/2$, the natural maximum-entropy ensemble is again the Haar ensemble.
We therefore define the trajectory-dependent distance from the Haar ensemble via the moments as 
\eq{
\Delta^{(k)}({\bf m}) = \frac{1}{2}\norm{\rho^{(k)}_{\rm PE}({\bf m})-\rho^{(k)}_{\rm Haar}}_1\,,
\label{eq:Deltak-m}
}
and study its average over trajectories and circuit realisations, $\braket{\Delta^{(k)}}=\mathbb{E}_{\cal C}\sum_{\bf m}p({\bf m})\Delta^{(k)}({\bf m})$.
In the opposite limit of $r\to 1^-$, since the hybrid dynamics purify $R$, it is effectively in a product state with $S$ such that $\ket{\Psi_{RS}({\bf m})}=\ket{\psi_R({\bf m})}\otimes\ket{\psi_S({\bf m})}$.
This naturally means that the state of $R$ is completely insensitive to the measurement outcomes $z_S$ and ${\cal E}_{\rm PE}({\bf m})$ collapses onto the single state $\ket{\psi_R({\bf m})}$ -- this is a \textit{trivial ensemble} with $\Delta^{(k)}({\bf m})=k/(k+1)$.

The measurement-induced purification transition therefore manifests itself in complementary ways in the two ensembles.
The two ensembles therefore probe the entanglement phases in a complementary fashion.
In the volume-law or mixed phase, $R$ remains maximally entangled with $S$ for almost all trajectories on timescales $t=O(N)$~\cite{Gullans2020Purification,gullans2020scalable}, such that ${\cal E}_{\rm TE}\to{\cal E}_{\mathbb{I}}$ and ${\cal D}^{(k)}_{\mathbb{I}}\to 0$.
The information of $R$ is, however, encoded non-locally in $S$, and the boundary measurements are expected to generate a Haar ensemble on $R$, $\Delta^{(k)}({\bf m})\to 0$.
In the area-law or pure phase, the bulk measurements purify $R$ which disentangles it from $S$. The TE then comprises the pure states $\ket{\psi_R({\bf m})}$, and the randomness of the bulk outcomes may instead lead to ${\cal E}_{\rm TE}$ approaching a maximum-entropy, pure state ensemble such as
the Haar ensemble, ${\cal D}^{(k)}_{\rm Haar}\to 0$.
The Haar ensemble therefore emerges in the TE in the pure phase, with the randomness supplied by the bulk outcomes, whereas it may do so in the PE in the mixed phase, with the randomness supplied by the boundary outcomes.

In the following,
we establish the ensemble transitions, as diagnosed by ${\cal D}^{(k)}_{\cal X}$ and $\braket{\Delta^{(k)}}$, in hybrid circuits 
of the form in Fig.~\ref{fig:setting}.
Specifically, we consider two families of brickwork circuits -- one comprised of Haar-random unitary gates~\cite{skinner2019measurement,li2019measurement,zabalo2020critical} and the other comprised of random Clifford gates with Pauli measurements ~\cite{li2018quantum,li2019measurement,Gullans2020Purification,gullans2020scalable,zabalo2020critical}. 

Clifford circuits are of particular interest in this regard, as they host an MIPT and are efficiently simulable classically for large $N$~\cite{gottesman1998heisenberg,aaronson2004improved}, but generate no non-stabiliserness, or magic~\cite{Veitch2014theresource,zhu2016clifford,leone2022stabilizer}.
This leads to ensembles generated in such cases resisting forming a design.
Specifically, we show that once the trajectories of different circuit realisations are pooled into a single ensemble, the TE in these circuits forms an exact state 3-design in the pure phase, but not a 4-design, owing to the absence of magic.
The PE, on the other hand, is restricted to a state 1-design for stabiliser states~\cite{loio2025quantum}, such that across the Clifford MIPT it only changes from a 1-design to the trivial ensemble.
This is in contrast to the Haar-random circuits where the TE and the PE do indeed form $k$-designs for arbitrary $k$ in the pure and mixed phases respectively.
This is a case in point that our results lead to a finer characterisation of the measurement-induced ensembles and transitions therein beyond the MIPT.

For completeness we also study all-to-all circuits with forced measurements~\cite{nahum2021measurement}, the results for which are relegated to Appendix~\ref{sec:sm-alltoall}.

\section{Ensemble transition in trajectories \label{sec:te}}

\begin{figure}[t]
    \centering
    \includegraphics[width=\linewidth]{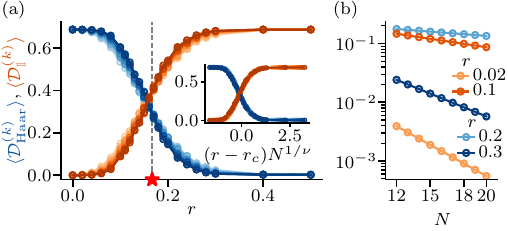}
    \caption{Ensemble transition in the TE in 1+1D brickwork circuits with Haar-random unitary gates and of depth $t=2N$.
    (a) The circuit-averaged distances $\braket{{\cal D}^{(k)}_{\rm Haar}}$ (blue tones) and $\braket{{\cal D}^{(k)}_{\mathbb{I}}}$ (orange tones) of the TE from the Haar and the identity ensembles, [defined in Eq.~\eqref{eq:Dk-te}], as a function of $r$ for $k=4$; darker shades denote larger $N$, with $N=12,13,\ldots,20$.
    The dashed line (star) marks $r_c\approx0.168$.
    Each distance saturates to $1-(k+1)/2^k=11/16$ (horizontal dashed line) when the TE coincides with the other reference ensemble.
    Inset: Scaling collapse of the data as a function of $(r-r_c)N^{1/\nu}$ with $r_c\approx0.168$ and $\nu\approx1.31$.
    (b) Exponential decay of the distances $\braket{{\cal D}^{(k)}_{\mathbb{I}}}$ (orange markers) and $\braket{{\cal D}^{(k)}_{\rm Haar}}$ (blue markers) with $N$ for $r<r_c$ and $r>r_c$ respectively.}
    \label{fig:Dkvsp}
\end{figure}

To probe the transition in the trajectories, we present results for the circuit averaged $\braket{{\cal D}^{(k)}_{\cal X}}\equiv\mathbb{E}_{\cal C}[{\cal D}^{(k)}_{\cal X}]$\footnote{Note that the distance measure ${\cal D}_{\cal X}^{(k)}$ is averaged over the circuits; the ensemble is over the trajectories for a given circuit. Since ${\cal D}_{\cal X}^{(k)}>0$ for every circuit, $\braket{{\cal D}_{\cal X}^{(k)}}\to 0$ implies that ${\cal D}_{\cal X}^{(k)}\to 0$ for every circuit.}
with ${\cal X}={\rm Haar},\mathbb{I}$ as a function of the measurement rate $r$.

\subsection{1+1D Haar-random circuits}

We start with the case of a 1+1D Haar-random circuit.
The results are shown in 
Fig.~\ref{fig:Dkvsp}(a)
for $k=4$ which shows a crossing of the data for different $N$ at a putative critical point $r_c$.
For $r<r_c$, $\braket{{\cal D}^{(k)}_{\mathbb{I}}}\to0$ and $\braket{{\cal D}^{(k)}_{\rm Haar}}\to1-(k+1)/2^k=11/16$ (for $k=4$).
This indicates that the TE is the identity ensemble, as $R$ is maximally mixed for almost all trajectories in the mixed phase.
For $r>r_c$, the roles are reversed and the TE approaches the Haar ensemble.
In both phases, the approach to the respective ensembles is exponential in $N$ [Fig.~\ref{fig:Dkvsp}(b)].
With the scaling ansatz $\braket{{\cal D}^{(k)}_{\cal X}}=f_{\cal X}[(r-r_c)N^{1/\nu}]$, the data for all $N$ collapses onto a single curve for $r_c\approx0.168$ and $\nu\approx1.31$ [inset of Fig.~\ref{fig:Dkvsp}(a)].
Both are in agreement with the critical properties of the MIPT, namely $r_c=0.168(5)$ and $\nu=1.2(2)$ obtained from the tripartite mutual information in the same model~\cite{zabalo2020critical}.
This establishes the presence of an ensemble transition in the trajectories.
It is important to note that 
the first moment of the TE is $\mathbb{I}/2$ for all $r$, whereas its higher moments change from those of the identity ensemble to those of the Haar ensemble across the transition.
This transition is therefore an example of a deep ergodicity-breaking transition wherein the ensemble is featureless at the level of its first moment for all $r$ but higher-moments reflect the presence of full Hilbert-space ergodicity or its absence across the transition.
In addition, note that in the pure phase, 
$R$ is purified to $\ket{\psi_R({\bf m})}$ for almost all trajectories.
However, purity alone places no constraint on how these states are distributed and hence, the emergence of the Haar ensemble in the pure phase for any fixed circuit driven by the Born-rule randomness of the bulk outcomes, is a non-trivial result. 
In fact, this need not always be the case as we shall show next in the context of Clifford circuits.

\begin{figure}[t]
    \centering
    \includegraphics[width=\linewidth]{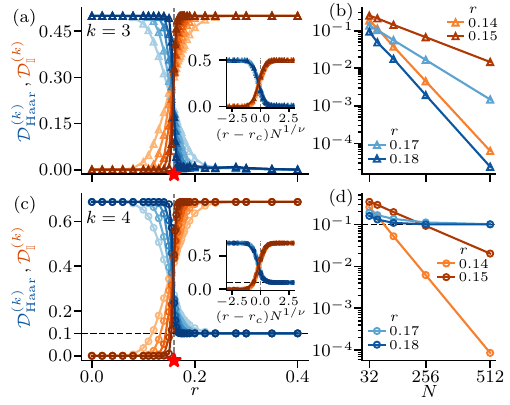}
    \caption{Ensemble transition in the TE pooled over bulk outcomes and circuit realisations, Eq.~\eqref{eq:te-clifford}, for 1+1D brickwork circuits with random Clifford gates. ${\cal D}^{(k)}_{\rm Haar}$ (blue, tones) and ${\cal D}^{(k)}_{\mathbb{I}}$ (orange,tones) as a function of $r$ for (a) $k=3$ and (c) $k=4$. Darker shades denote larger $N=32,64,128,256,512$.
    The dashed line (star) marks $r_c\approx0.16$. For $r<r_c$, ${\cal D}^{(k)}_{\mathbb{I}}$ vanishes exponentially with $N$ for both $k=3$ and $4$ [see (b) and (d) for scaling with $N$]. For $r>r_c$, ${\cal D}^{(k=3)}_{\rm Haar}$ vanishes exponentially with $N$ but ${\cal D}^{(k=4)}_{\rm Haar}$ saturates to a finite value indicating the formation of only a 3-design.
    Insets in (a) and (c): Scaling collapse of the data as a function of $(r-r_c)N^{1/\nu}$ with $r_c\approx0.16$ and $\nu\approx1.3$.
    }
    \label{fig:cliffordTE}
\end{figure}

\subsection{1+1D Clifford circuits and 3-designs}
We next turn to 1+1D brickwork circuits with random Clifford unitaries and Pauli-$Z$ measurements. 
For reasons which will shortly become clear, we make the dependence of the ensemble on the circuit realisation, ${\cal C}$ explicit, 
and denote by $\rho_R({\bf m}|{\cal C})$ the state of $R$ for the record of outcomes ${\bf m}$ in ${\cal C}$.
The key point about the resulting stabiliser states is that $\rho_R({\bf m}|{\cal C})$
is either $\mathbb{I}/2$ or one of the six pure stabiliser states $\Phi_{a,s}=(\mathbb{I}+s\sigma^a)/2$, with $a\in\{x,y,z\}$ and $s=\pm 1$~\cite{gullans2020scalable}.
Whether $R$ purifies, and along which axis $a$, is fixed by ${\cal C}$, and the bulk outcomes only decide the sign $s$.
The TE of a single ${\cal C}$ is therefore either the identity ensemble with $\rho_R({\bf m}|{\cal C})=\mathbb{I}/2~\forall {\bf m}$ or an ensemble with exactly two pure states, $\Phi_{a,\pm}$.
Crucially, neither case forms a 2-design, in contrast to the Haar circuits, where the TE for a typical ${\cal C}$ is the Haar ensemble in the pure phase.
This presents a point of departure for the ensembles in Clifford circuits from those in Haar-random circuits despite both hosting MIPTs.

To see if the randomness arising from the circuit realisations in addition to the measurement outcomes can lead to higher designs, we generate the ensemble over both ${\cal C}$ and ${\bf m}$
by pooling the trajectories of all circuit realisations into a single ensemble, in which each trajectory is labelled by both ${\cal C}$ and ${\bf m}$,
\eq{
{\cal E}_{\rm TE}&=\{p({\cal C})\,p({\bf m}|{\cal C}),\rho_R({\bf m}|{\cal C})\}_{{\cal C},{\bf m}}\,,}
with the moments given by 
\eq{
\braket{\rho^{(k)}_{\rm TE}}&=\mathbb{E}_{\cal C}\left[\sum_{\bf m}p({\bf m}|{\cal C})\,\rho_R({\bf m}|{\cal C})^{\otimes k}\right]\,,
\label{eq:te-clifford}
}
and compute ${\cal D}^{(k)}_{\cal X}$ from Eq.~\eqref{eq:Dk-te} with $\rho^{(k)}_{\rm TE}\to\braket{\rho^{(k)}_{\rm TE}}$.
In contrast to $\mathbb{E}_{\cal C}[{\cal D}_{\cal X}^{(k)}]$, the circuit average now enters the moments, and by the convexity of the trace norm the resulting distances never exceed $\braket{{\cal D}^{(k)}_{\cal X}}$.

We summarise the structure of this ensemble below and present the derivations in Appendix~\ref{sec:sm-te}.
Using the fact that the first gate acting on the Bell-pair partner of $R$ is a random Clifford gate, courtesy $U_\tau$, all six stabiliser states are equally likely such that
\eq{
\braket{\rho^{(k)}_{\rm TE}}=q\left(\frac{\mathbb{I}}{2}\right)^{\otimes k}+\frac{1-q}{6}\sum_{a,s}\Phi_{a,s}^{\otimes k}\,,
\label{eq:te-clifford-result}
}
where $q$ is the probability that the $\rho_{R}({\bf m}|{\cal C})$ remains mixed, which is simply the order parameter for the purification transition of $R$ for Clifford circuits~\cite{gullans2020scalable}.
The six states, $\{\Phi_{a,s}\}$, form an octahedron on the Bloch sphere, which reproduces the first three moments of the Haar ensemble but not the fourth~\cite{kueng2015qubit,zhu2016clifford}. This implies from Eq.~\eqref{eq:te-clifford-result} that the $k^{\rm th}$ moment of the TE for $k\leq 3$ is mixture of the $k^{\rm th}$ moment of ${\cal E}_{\mathbb{I}}$ with weight $q$ and that of the ${\cal E}_{\rm Haar}$ with weight $1-q$. 
Therefore the distances for $k\leq 3$ are given by
\eq{
{\cal D}^{(k)}_{\rm Haar}\!=q\!\!\left[1-\frac{k+1}{2^k}\right]\!\!,~{\cal D}^{(k)}_{\mathbb{I}}\!=(1-q)\!\!\left[1-\frac{k+1}{2^k}\right].
\label{eq:Dk-clifford}
}
On the other hand, from Eq.~\eqref{eq:te-clifford-result} for $k=4$ we have
\eq{
{\cal D}^{(4)}_{\rm Haar}\!\!=\max\left(\frac{1}{10}+\frac{5q}{16},\frac{11q}{16}\right)\!,~{\cal D}^{(4)}_{\mathbb{I}}=\frac{11(1-q)}{16}.
\label{eq:Dk-clifford-4}
}
The TE thus goes from the identity ensemble at $q=1$ in the mixed phase to an exact state 3-design at $q=0$ in the pure phase, which remains at a distance $1/10$ from the Haar ensemble for $k=4$.
Numerical results in Fig.~\ref{fig:cliffordTE} confirm this for $k=3$ and $k=4$.
As the distances depend on $q$ alone, the data collapses with $r_c\approx0.16$ and $\nu\approx1.3$, in agreement with $r_c=0.1598(5)$ and $\nu=1.30(5)$ of the Clifford MIPT~\cite{gullans2020scalable}.
These results show that augmenting the TE with the randomness of the circuit realisation leads to an ensemble transition in Clifford circuits from the identity ensemble to a 3-design. 

The distance from the Haar ensemble at $k=4$ is set by the absence of magic.
For a pure state with Bloch vector ${\bf r}$, ${\rm Tr}[\rho^{\otimes4}\sum_a(\sigma^a)^{\otimes4}]=\sum_ar_a^4=2^{1-M_2}-1$, where $M_2$ is the stabiliser 2-R\'enyi entropy~\cite{leone2022stabilizer}.
This equals $1$ for every stabiliser state and $3/5$ on average over the Haar ensemble, such that without magic the TE cannot be a $k$-design for any $k\geq4$, however random the circuits and the outcomes.
The Haar circuits of Fig.~\ref{fig:Dkvsp} close this gap as their gates generate magic.

\section{Deep thermalisation transition in the PE \label{sec:pe}}
\begin{figure}[t]
    \centering
    \includegraphics[width=\linewidth]{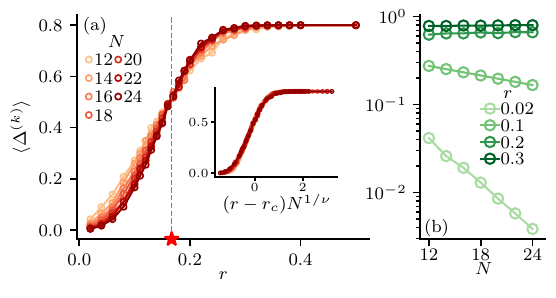}
    \caption{Deep thermalisation of the PE across the purification transition for the case of Haar-random circuits.
    (a) The trajectory-averaged distance $\braket{\Delta^{(k)}}$ of the PE on $R$ from the Haar ensemble, Eq.~\eqref{eq:Deltak-m}, as a function of the measurement rate $r$ for $k=4$ and different $N$.
    The dashed line (star) marks $r_c\approx0.168$.
    For $r>r_c$, the data saturates to $k/(k+1)=4/5$, the value for the trivial ensemble.
    Inset: Scaling collapse of the data as a function of $(r-r_c)N^{1/\nu}$ with $r_c\approx0.168$ and $\nu\approx1.31$.
    (b) Exponential decay of $\braket{\Delta^{(k)}}$ with $N$ for $r<r_c$ and its saturation to $k/(k+1)=4/5$ for $r>r_c$.}
    \label{fig:deltakvsp}
\end{figure}

We next
address the fate of the PE, Eq.~\eqref{eq:pe-def}, for a given circuit ${\cal C}$ and bulk measurement outcomes ${\bf m}$ as a function of $r$ as diagnosed by $\braket{\Delta^{(k)}}$ defined in Eq.~\eqref{eq:Deltak-m}.
The results for the Haar-random circuits are shown in Fig.~\ref{fig:deltakvsp} for $k=4$.
As a function of $r$, $\braket{\Delta^{(k)}}$ shows a crossing for different $N$ at $r_c$. 
For $r<r_c$, $\braket{\Delta^{(k)}}\to 0$ exponentially with $N$ whereas for $r>r_c$, it saturates with $N$ to $k/(k+1)=4/5$.
This indicates that the PE approaches a Haar ensemble for $r<r_c$ whereas for $r>r_c$ it is a trivial ensemble with just one state. 
Note that since $\Delta^{(k)}({\bf m})\ge 0$, the fact that $\braket{\Delta^{(k)}}\to 0$ for $r<r_c$ implies that the PE forms a Haar ensemble for typical circuits and bulk measurement sequences. 
In other words, $R$ thermalises deeply for typical trajectories, 
and the randomness of the boundary outcomes $z_S$ suffices to generate an approximate state $k$-design on $R$ as it does under unitary dynamics~\cite{cotler2023emergent,ho2022exact,ippoliti2022solvablemodelofdeep,mark2024maximum}. This is consistent with hybrid circuits in the mixed phase exhibiting properties of the unitary circuits for exponentially long, in $N$, timescales.
For $r>r_c$, on the other hand, the bulk measurements purify $R$ well within the $O(N)$ depth, following which the state of $R$ is agnostic to the boundary measurement outcomes $z_S$, and the PE is therefore a trivial ensemble.

Under the same scaling ansatz as for the TE, the data for all $N$ collapses onto a single curve for $r_c\approx0.168$ and $\nu\approx1.31$ [inset of Fig.~\ref{fig:deltakvsp}(a)], such that the PE and the TE undergo their ensemble transitions together, at the MIPT.
To establish the generality of these results, we show in Appendix~\ref{sec:sm-alltoall} that both ensemble transitions also occur across the forced-measurement transition in all-to-all circuits~\cite{nahum2021measurement}, where they coincide with the entanglement transition and inherit its Kosterlitz-Thouless-like scaling.

\section{Ensemble transitions and the MIPT \label{sec:mipt}}
The MIPT in the trajectories is a manifestation of the underlying measurement-induced ensemble transitions in both the TE and the PE. 
This connection can be understood more concretely from a bound relating the moments of the PE to those of the TE.
Labelling the $k$ copies of $R$ in Eq.~\eqref{eq:pe-moments} as $R_1,\cdots,R_k$ and tracing out all but the first copy from the $k^{\rm th}$ moment reduces it to the first moment
\eq{
{\rm Tr}_{R_2\cdots R_k}\,\rho^{(k)}_{\rm PE}({\bf m})=\rho_R({\bf m})\,.
}
Since the trace distance does not increase under partial traces, tracing out all but one of the $k$ copies in Eq.~\eqref{eq:Deltak-m} leads to the inequality 
\eq{
\braket{\Delta^{(k)}}\geq\frac{1}{2}\,\Braket{\norm*{\rho_R({\bf m})-\frac{\mathbb{I}}{2}}_1}\,.
\label{eq:Deltak-bound}
}
Given that $R$ is a single qubit, its state can be generally written as $\rho_R({\bf m}) = [\mathbb{I}+{\mathbf{r}}(\mathbf{m})\cdot{\bm \sigma}]/2$ such that the right hand side of Eq.~\eqref{eq:Deltak-bound} is simply $\braket{{\bf |r|}}/2$ which is straightforwardly related to the purity of $R$ averaged over the TE -- this is precisely the order parameter for the purification transition~\cite{Gullans2020Purification,gullans2020scalable}.
It vanishes only if $R$ remains maximally mixed in almost all trajectories and the TE forms an identity ensemble.
This therefore guarantees that the PE can thermalise deeply only for $r<r_c$ when the TE forms an identity ensemble.

However, the bound only presents a necessary and not a sufficient condition for deep thermalisation of the PE to the Haar ensemble. 
A maximally mixed $\rho_R({\bf m})$ only fixes the first moment of ${\cal E}_{\rm PE}({\bf m})$ to identity and does not mandate anything for the higher moments. 
Deep thermalisation of the PE, indicated by $\braket{\Delta^{(k)}}\to0$ requires in addition that the boundary outcomes $z_S$ distribute the states $\ket{\psi_R({\bf m},z_S)}$ uniformly over the Bloch sphere.
That the PE thermalises deeply throughout the mixed phase in the Haar-random circuits, with the same $r_c$ and $\nu$ as the MIPT, indicates that for Haar-random gates this is automatic: as long as $R$ remains entangled with $S$, the unitaries scramble its information sufficiently, and the ensemble transition is slaved to the purification transition. As we discuss shortly, this is not the case in Clifford circuits.
Before doing so, we make some comments about a natural interpretation of our results 
in terms of entanglement as a resource for deep thermalisation.

For the boundary measurements to generate a non-trivial PE on $R$, $R$ must be entangled with $S$, and Eq.~\eqref{eq:Deltak-bound} makes this quantitative, as any loss of entanglement between $R$ and $S$ bounds $\braket{\Delta^{(k)}}$ away from zero.
In the hybrid circuit, the unitary gates generate this resource and spread it across $S$, whereas the bulk measurements destroy it.
In the mixed phase, loosely speaking, the unitaries dominate the measurements, $R$ remains maximally entangled with $S$, and deep thermalisation survives; in the pure phase, the measurements consume the resource entirely and deep thermalisation is lost.
The MIPT therefore offers a direct route to studying the role of entanglement as a resource in deep thermalisation, with the bulk measurement rate $r$ acting as a knob that controls the amount of resource available to the boundary measurements.
This complements recent works in which coherence and non-stabiliserness were identified as the resources underpinning deep thermalisation, and the corresponding transitions were tied to whether these resources can be localised onto the unmeasured subsystem~\cite{liu2026coherence,feng2026quantum,varikuti2025deep}.

Returning to Clifford circuits, they show that entanglement alone does not suffice for the PE to thermalise deeply.
As $\ket{\Psi_{RS}({\bf m})}$ is a stabiliser state, even when $R$ is maximally entangled with $S$, measuring $S$ leaves $R$ in one of the two eigenstates of a single Pauli operator with equal probabilities.
${\cal E}_{\rm PE}({\bf m})$ is therefore only a 1-design, with $\Delta^{(k)}({\bf m})=(k-1)/(k+1)$, as is the case for the PE of any stabiliser state~\cite{loio2025quantum}.
Across the Clifford MIPT, the PE thus changes from a 1-design to the trivial ensemble. Higher designs require magic, which, in case of the Haar-random brickwork circuits is supplied by the non-stabiliserness of the gates.
The PE and the TE (defined over both bulk measurements and circuits) therefore require magic at different orders to thermalise deeply: the TE only beyond the third moment, as the random gate acting on the Bell partner of $R$ randomises the axis that the bulk measurements read out, and the PE already beyond the first.

\section{Summary and outlook \label{sec:summary}}
In this work, we showed that the two ensembles of states natural to hybrid quantum circuits, the TE defined over trajectories of measurement outcomes and the PE obtained by measuring all but a few qubits given a trajectory, undergo ensemble phase transitions that coincide with the MIPTs. Crucially, these transitions are invisible to the averaged dynamics. For Haar-random circuits, the TE transitions from the identity ensemble in the mixed phase to the Haar ensemble in the pure phase, while the PE transitions from the Haar ensemble to the trivial ensemble. By establishing a bound on the distance between the PE and the Haar ensemble in terms of the purity averaged over the TE [see Eq.~\eqref{eq:Deltak-bound}], we explicitly tied the two ensembles together, cementing entanglement as a necessary resource for the deep thermalisation of the PE. However, our analysis of hybrid Clifford circuits reveals that entanglement alone is not sufficient. Despite hosting an MIPT, the Clifford TE reaches only a state 3-design (and only when pooled over circuits), while the PE degrades from a 1-design to the trivial ensemble. Consequently, statistical ensembles of states provide a strictly finer characterisation of measurement-induced phases, distinguishing circuits which otherwise appear identical from the point of view of entanglement phases.

Looking forward, while we focused on the ensembles of a reference qubit initially entangled with the system, conventional MIPT settings without ancillas offer a rich landscape for further ensemble transitions. Starting from a mixed state in such models, each trajectory in the mixed phase is expected to yield a mixed state -- though not necessarily maximally mixed, as was the case here -- and a pure state in the pure phase. It remains an open question whether the TE in this setting undergoes a phase transition from a mixed-state deep-thermalised ensemble to a pure-state deep-thermalised ensemble across the MIPT. Similarly, starting from a pure state, the fate of the PE generated on a small subsystem warrants investigation. Specifically, it will be important to determine if the deep thermalisation of the PE, well-established for unitary dynamics, survives in the low-measurement-rate phase; that is, whether the states retain sufficient entanglement for the PE to deep thermalise and if they do, how to determine the appropriate Scrooge ensemble.

This naturally connects to the broader question of constructing resource-constrained, maximum-entropy ensembles in the context of mixed-state deep thermalisation~\cite{yu2025mixed,sherry2026deep,mandal2025partialprojectedensemblesspatiotemporal,yu2026mixed}. In the pure phase of hybrid circuits, the TE realizes a maximum-entropy pure-state ensemble constrained both by its first moment and its underlying resource content. This constraint is sharply reflected in the contrast between Haar-random and Clifford circuits; the latter restricts the TE to a 3-design due to the strict absence of non-stabiliserness. Exploring this framework in the mixed phase where the global state of a trajectory remains mixed, but not maximally so, presents an exciting avenue for identifying analogous resource-constrained, maximum-entropy {\it mixed-state} ensembles generated by hybrid dynamics.

Finally, pushing further into the role of quantum resources, an inherently interesting direction is whether measurements can also control the amount of magic in the ensembles. A natural testbed for this is hybrid Clifford circuits punctuated by non-Clifford measurements. Here, the unitary gates remain Clifford, and magic is injected exclusively through measurements in a non-stabiliser basis. Under these conditions, the states of $R$ are no longer confined to the seven states underlying Eq.~\eqref{eq:te-clifford-result}, allowing for the potential emergence of higher designs. The sharp question then becomes whether the TE and PE can surpass a 3-design and 1-design, respectively, and if this holds even for the TE of a single circuit realization. This is intimately connected to magic transitions in monitored circuits~\cite{niroula2024phase,bejan2024dynamical,fux2024entanglement,tarabunga2025magic}, which can decouple from the standard entanglement transition~\cite{bejan2024dynamical,fux2024entanglement}. Because the PE's deviation from a $k$-design is fundamentally linked to the stabiliser entropies of the pre-measurement state~\cite{loio2025quantum}, a magic transition may naturally manifest as a transition in the design order. This raises the intriguing possibility of a hierarchy of ensemble transitions driven by $k$~\cite{mandal2026upcoming} that occur deep within the mixed phase where sufficient entanglement is guaranteed by Eq.~\eqref{eq:Deltak-bound}, but the design order is strictly dictated by the magic. Selectively measuring only $S$ at the final time in a non-stabilizer basis, or restricting such measurements to the bulk, could further disentangle the distinct roles magic plays in the PE versus the TE.

\begin{acknowledgements} The authors thank A. Sherry, R. Thakur, and K. Sorari for useful discussions. This work was supported by the Department of Atomic Energy, Government of India, under project nos. RTI4019 and RTI4013.
S.R. acknowledges support from SERB-DST (India) under Grant No. SRG/2023/000858, from ANRF (India) under Grant No. ANRF/ARG/2025/004045/PS, and from a Max Planck Partner Group grant between ICTS-TIFR, Bengaluru and MPIPKS, Dresden.
\end{acknowledgements}

\appendix

\section{Trajectory ensemble in hybrid Clifford circuits\label{sec:sm-te}}

In this section, we derive the results for the TE in hybrid Clifford circuits quoted in the main text.
Specifically, we show that the moments of the TE pooled over bulk outcomes and circuit realisations, Eq.~\eqref{eq:te-clifford}, is given by Eq.~\eqref{eq:te-clifford-result}, with $q$ the order parameter of Ref.~\cite{gullans2020scalable}, and that this leads to the results for the distances in Eqs.~\eqref{eq:Dk-clifford} and \eqref{eq:Dk-clifford-4}.
The TE is therefore an exact state 3-design for $q=0$, but not a $k$-design for any $k\geq4$.
Throughout, we use the shorthand
\eq{
&H_k\equiv\rho^{(k)}_{\rm Haar}=\frac{\Pi^{(k)}_{\rm sym}}{k+1}\,,~~
I_k\equiv\rho^{(k)}_{\mathbb{I}}=\left(\frac{\mathbb{I}}{2}\right)^{\otimes k}\,,
}
for the moments of the two reference ensembles in Eq.~\eqref{eq:Dk-te},
and
\eq{
\Phi_{a,s}=\frac{\mathbb{I}+s\sigma^a}{2}\,,
\label{eq:sm-notation}
}
with $a\in\{x,y,z\}$ and $s=\pm1$, for the six pure stabiliser states of $R$, which we refer to as poles as they lie at the ends of the three axes of the Bloch sphere.

\subsection{Stabiliser states and their dynamics}
A pure stabiliser state $\ket{\Psi}$ of $N+1$ qubits is the unique state with $g\ket{\Psi}=\ket{\Psi}$ for all $g$ in an abelian group $G$ of $2^{N+1}$ Pauli strings, with $-\mathbb{I}\notin G$~\cite{gottesman1998heisenberg,aaronson2004improved}.
Its projector is
\eq{
\ket{\Psi}\bra{\Psi}=\frac{1}{2^{N+1}}\sum_{g\in G}g\,.
\label{eq:sm-stab}
}
The initial state $\ket{\Phi^+}_{RS}$ in Eq.~\eqref{eq:psi-traj} is one such state, with $G$ generated by $\sigma^x_R\sigma^x_1$, $\sigma^z_R\sigma^z_1$ and $\sigma^z_i$ for $i\geq2$.
A Clifford gate $U$ maps $G\mapsto UGU^\dagger$, which is again a group of Pauli strings.
A measurement of a Pauli string $P$, here $\sigma^z_i$, falls into one of two cases.
If $P$ commutes with all of $G$, then $P$ or $-P$ belongs to $G$, such that the outcome is deterministic and the state is unchanged.
Otherwise, there exists a $g\in G$ that anticommutes with $P$, such that $\braket{\Psi|P|\Psi}=\braket{\Psi|Pg|\Psi}=-\braket{\Psi|gP|\Psi}=-\braket{\Psi|P|\Psi}=0$.
The two outcomes $s=\pm1$ then occur with probability $1/2$ each, and the post-measurement state is stabilised by $sP$ together with the elements of $G$ that commute with $P$.
Hence, $\ket{\Psi_{RS}({\bf m})}$ in Eq.~\eqref{eq:psi-traj} is a stabiliser state along every trajectory.

\subsection{Seven possible states of $R$} Consider an element $g=g_R\otimes g_S$ of $G$.
As Pauli strings other than the identity are traceless, ${\rm Tr}_S[g_R\otimes g_S]$ vanishes unless $g_S=\mathbb{I}_S$, in which case it equals $2^Ng_R$.
Tracing out $S$ in Eq.~\eqref{eq:sm-stab} therefore retains only the subgroup $G_R$ of stabilisers that act trivially on $S$,
\eq{
\rho_R=\frac{1}{2}\sum_{g_R\in G_R}g_R\,.
\label{eq:sm-rhoR}
}
Distinct nontrivial Pauli matrices anticommute whereas elements of $G$ commute, so at most one of them can appear in $G_R$, and $-\mathbb{I}\notin G$ rules out it appearing with both signs.
$G_R$ thus contains either the identity alone or the identity and a single $s\sigma^a$, which from Eq.~\eqref{eq:sm-rhoR} gives $\rho_R=\mathbb{I}/2$ or $\rho_R=\Phi_{a,s}$ respectively.
In other words, $R$ is either maximally entangled with $S$ or disentangled from it, and its von Neumann entropy in bits, $S[\rho_R]=-{\rm Tr}\,\rho_R\log_2\rho_R$, takes only the values $1$ and $0$.

\subsection{A single circuit realisation} We now fix ${\cal C}$, namely the gates and the measurement locations, and compare different records of outcomes.
Any two allowed records ${\bf m}$ and ${\bf m}'$ lead to states related by a Pauli string $Q$,
\eq{
\ket{\Psi_{RS}({\bf m}')}\propto Q\ket{\Psi_{RS}({\bf m})}\,,
\label{eq:sm-frame}
}
which we show by induction over the operations in the circuit.
Eq.~\eqref{eq:sm-frame} holds at the start with $Q=\mathbb{I}$.
A Clifford gate $U$ preserves it, as $UQ\ket{\Psi}=(UQU^\dagger)U\ket{\Psi}$ and $UQU^\dagger$ is a Pauli string.
For a measurement of $P$, the stabiliser groups of $\ket{\Psi}$ and $Q\ket{\Psi}$ differ only in the signs of their elements, so that the two states fall in the same case above.
In the deterministic case, both states are unchanged.
In the random case, moving $Q$ through the projector gives $(\mathbb{I}+s'P)Q\ket{\Psi}=Q(\mathbb{I}+s''P)\ket{\Psi}$, with $s''=\pm s'$ depending on whether $Q$ and $P$ commute or anticommute.
Let $s$ be the outcome on $\ket{\Psi}$.
If $s''=s$, Eq.~\eqref{eq:sm-frame} holds with the same $Q$.
If $s''=-s$, the stabiliser $g$ that anticommutes with $P$ gives $(\mathbb{I}-sP)\ket{\Psi}=(\mathbb{I}-sP)g\ket{\Psi}=g(\mathbb{I}+sP)\ket{\Psi}$, and Eq.~\eqref{eq:sm-frame} holds with $Q\mapsto Qg$.

Eq.~\eqref{eq:sm-frame} has two consequences.
First, the number of random measurements is the same for all records, such that all allowed records of a given ${\cal C}$ are equally likely.
Second, writing $Q=Q_R\otimes Q_S$ gives $\rho_R({\bf m}'|{\cal C})=Q_R\,\rho_R({\bf m}|{\cal C})\,Q_R$, where the Pauli matrix $Q_R$ leaves $\mathbb{I}/2$ invariant and maps $\Phi_{a,s}\mapsto\Phi_{a,\pm s}$.
Whether $R$ purifies, and along which axis $a_{\cal C}$, is therefore the same for all records of a given ${\cal C}$, and the bulk outcomes only decide the sign $s$.

The two signs are in fact equally likely.
Since $K_{\bf m}$ in Eq.~\eqref{eq:psi-traj} acts only on $S$ and $\sum_{\bf m}K_{\bf m}^\dagger K_{\bf m}=\mathbb{I}$, the first moment of the TE is that of the initial state,
\eq{
\sum_{\bf m}p({\bf m}|{\cal C})\,\rho_R({\bf m}|{\cal C})={\rm Tr}_S\left[U_\tau\ket{\Phi^+}\bra{\Phi^+}_{RS}U_\tau^\dagger\right]=\frac{\mathbb{I}}{2}\,,
\label{eq:sm-first-moment}
}
which, if $R$ purifies along $a_{\cal C}$, is only possible if $\Phi_{a_{\cal C},+}$ and $\Phi_{a_{\cal C},-}$ have equal weights.
With $q_{\cal C}=1$ if $R$ remains mixed in ${\cal C}$ and $q_{\cal C}=0$ otherwise, the moments of the TE of a single ${\cal C}$ are therefore
\eq{
\rho^{(k)}_{\rm TE}({\cal C})=q_{\cal C}\,I_k+\frac{1-q_{\cal C}}{2}\sum_{s}\Phi_{a_{\cal C},s}^{\otimes k}\,.
\label{eq:sm-te-fixed}
}
A single ${\cal C}$ thus selects a single axis, which is not sufficient for a 2-design.
For $k=2$, the second term of Eq.~\eqref{eq:sm-te-fixed} involves $\frac{1}{2}\sum_s\Phi_{a,s}^{\otimes2}=\frac{1}{4}(\mathbb{I}\otimes\mathbb{I}+\sigma^a\otimes\sigma^a)$, whereas
\eq{
H_2=\frac{1}{4}\mathbb{I}\otimes\mathbb{I}+\frac{1}{12}\sum_{b}\sigma^b\otimes\sigma^b\,.
\label{eq:sm-H2}
}
Their difference has eigenvalues $1/6$, $1/6$, $-1/3$ and $0$, such that ${\cal D}^{(2)}_{\rm Haar}=1/3$ for every ${\cal C}$ in which $R$ purifies, for any $N$.
This is in contrast to the Haar circuits of Fig.~\ref{fig:Dkvsp}, and it is why the circuit realisations need to be pooled in Eq.~\eqref{eq:te-clifford} for the TE to go beyond a 1-design.

\subsection{Purification order parameter} The order parameter of Ref.~\cite{gullans2020scalable} is the average entropy of $R$ over the records,
\eq{
S_Q({\cal C})=\sum_{\bf m}p({\bf m}|{\cal C})\,S[\rho_R({\bf m}|{\cal C})]\,,
\label{eq:sm-SQ}
}
further averaged over circuit realisations.
As $S[\rho_R({\bf m}|{\cal C})]$ is $1$ if $R$ is mixed and $0$ otherwise, $S_Q({\cal C})=q_{\cal C}$ circuit by circuit.
In the ensemble of Eq.~\eqref{eq:te-clifford}, the pair $({\cal C},{\bf m})$ occurs with the joint probability $p({\cal C})\,p({\bf m}|{\cal C})$.
The weight of $\mathbb{I}/2$ in this ensemble is therefore the total probability that $R$ remains mixed,
\eq{
q&\equiv\sum_{\cal C}p({\cal C})\sum_{\bf m}p({\bf m}|{\cal C})\,\delta_{\rho_R({\bf m}|{\cal C}),\mathbb{I}/2}\nonumber\\
&=\mathbb{E}_{\cal C}\,q_{\cal C}=\mathbb{E}_{\cal C}\,S_Q({\cal C})\,,
\label{eq:sm-q}
}
namely the fraction of circuit realisations in which $R$ remains mixed, which is the $q$ appearing in Eq.~\eqref{eq:te-clifford-result}.
Eq.~\eqref{eq:sm-q} holds with the entropy measured in bits; with the natural logarithm, $q=\mathbb{E}_{\cal C}\,S_Q({\cal C})/\ln2$.
It relies on the entropy of $R$ being either $0$ or $1$, and does not hold for Haar-random gates, for which $S[\rho_R({\bf m}|{\cal C})]$ takes any value in $[0,1]$.

\subsection{Circuit averaging democratises the poles} From the seven possible states of $R$, the moments in Eq.~\eqref{eq:te-clifford} have the form
\eq{
\braket{\rho^{(k)}_{\rm TE}}=q\,I_k+\sum_{a,s}\pi_{a,s}\,\Phi_{a,s}^{\otimes k}\,,
\label{eq:sm-te-pi}
}
where $\pi_{a,s}$ is the probability of the pole $\Phi_{a,s}$ in the pooled ensemble, with $\sum_{a,s}\pi_{a,s}=1-q$.
To fix the $\pi_{a,s}$, note that for any single-qubit unitary $V$, $(V\otimes\mathbb{I})\ket{\Phi^+}=(\mathbb{I}\otimes V^{\rm T})\ket{\Phi^+}$ for the Bell pair of $R$ and the qubit $i=1$ of $S$.
The first operation acting on this qubit is a two-qubit gate $G_1$ of $U_\tau$, drawn uniformly from the Clifford group.
Consider the realisation ${\cal C}'$ obtained from ${\cal C}$ by replacing $G_1\mapsto G_1(V^{\rm T}\otimes\mathbb{I})$, with $V$ a single-qubit Clifford, in which case $V^{\rm T}$ is also a Clifford.
${\cal C}'$ is as likely as ${\cal C}$, by the invariance of the uniform measure on the Clifford group.
Moreover, the extra $V^{\rm T}$ on the qubit $i=1$ is equivalent to $V$ on $R$, which commutes with the entire evolution as nothing else acts on $R$.
Every trajectory of ${\cal C}'$ is therefore that of ${\cal C}$ with $\ket{\Psi_{RS}({\bf m})}\mapsto V_R\ket{\Psi_{RS}({\bf m})}$ and the same Born probability, such that $\rho_R\mapsto V\rho_RV^\dagger$ and $q_{\cal C}$ is unchanged.

Single-qubit Cliffords act on the Bloch sphere as the 24 rotations of the octahedron, which map any pole $\Phi_{a,s}$ to any other.
Hence, $\pi_{a,s}=(1-q)/6$ for all six poles, and Eq.~\eqref{eq:sm-te-pi} becomes
\eq{
\braket{\rho^{(k)}_{\rm TE}}=q\,I_k+(1-q)\,S_k\,,~~S_k\equiv\frac{1}{6}\sum_{a,s}\Phi_{a,s}^{\otimes k}\,,
\label{eq:sm-te-avg}
}
at every time $t$, which is Eq.~\eqref{eq:te-clifford-result}.
Only this one gate needs to be random for the poles to be equalised.
If, instead, the qubit $i=1$ could be measured before any gate acts on it, $R$ would purify along the $z$ axis with probability $r$, and in the pure phase, Eq.~\eqref{eq:sm-H2} would give ${\cal D}^{(2)}_{\rm Haar}=r/3$.

\subsection{Formation of a 3-design but not a 4-design} For the states $\rho({\bf r})=\frac{1}{2}(\mathbb{I}+{\bf r}\cdot{\bm\sigma})$ with $|{\bf r}|=1$, the $k$-fold tensor power expands as
\eq{
\rho({\bf r})^{\otimes k}=\frac{1}{2^k}\sum_{A\subseteq\{1,\ldots,k\}}\prod_{i\in A}{\bf r}\cdot{\bm\sigma}_i\,,
\label{eq:sm-expand}
}
where ${\bm\sigma}_i$ acts on the $i^{\rm th}$ copy.
Each term is a monomial of degree $|A|\leq k$ in the components of ${\bf r}$, such that the $k^{\rm th}$ moment of any ensemble of pure states is fixed by the averages of these monomials.
$H_k$ is the average of Eq.~\eqref{eq:sm-expand} with ${\bf r}$ uniform on the sphere, and $S_k$ the average over the six poles, ${\bf r}\in\{\pm\hat x,\pm\hat y,\pm\hat z\}$.
Both sets are invariant under ${\bf r}\to-{\bf r}$, so that the odd moments vanish in both, and both have $\mathbb{E}\,r_ar_b=\delta_{ab}/3$.
Hence, $S_k=H_k$ for $k\leq3$~\cite{kueng2015qubit,webb2016clifford,zhu2017multiqubit}, and Eq.~\eqref{eq:sm-te-avg} gives
\eq{
\braket{\rho^{(k)}_{\rm TE}}-H_k&=q(I_k-H_k)\,,\nonumber\\
\braket{\rho^{(k)}_{\rm TE}}-I_k&=(1-q)(H_k-I_k)\,;~~k\leq3\,.
\label{eq:sm-diff3}
}
For a traceless Hermitian operator, half the trace norm is the sum of its positive eigenvalues.
$I_k-H_k$ has the eigenvalue $2^{-k}-1/(k+1)<0$ on the $(k+1)$-dimensional symmetric subspace and the eigenvalue $2^{-k}$ on its $(2^k-k-1)$-dimensional complement, such that $\frac{1}{2}\norm{I_k-H_k}_1=1-(k+1)/2^k$.
Eq.~\eqref{eq:sm-diff3} then gives
\eq{
{\cal D}^{(k)}_{\rm Haar}&=q\left[1-\frac{k+1}{2^k}\right]\,,\nonumber\\
{\cal D}^{(k)}_{\mathbb{I}}&=(1-q)\left[1-\frac{k+1}{2^k}\right]\,;~~k\leq3\,,
\label{eq:sm-Dk3}
}
which is Eq.~\eqref{eq:Dk-clifford}, and the TE is an exact 3-design for $q=0$.

We now turn to $k=4$, where the two ensembles first differ.
In the expansion in Eq.~\eqref{eq:sm-expand}, the moments up to degree three agree as shown above, whereas at degree four, the poles give $\mathbb{E}\,r_a^4=1/3$ and $\mathbb{E}\,r_a^2r_b^2=0$ for $a\neq b$, and the sphere gives $1/5$ and $1/15$.
Hence, $S_4\neq H_4$, as is known for stabiliser states in general~\cite{kueng2015qubit,zhu2016clifford}.
To quantify the difference, we compute $S_4$ explicitly.
As $\Phi_{a,s}^{\otimes4}$ is invariant under permutations of the copies, $S_4$ acts only within the five-dimensional symmetric subspace, as does $H_4$.
A convenient basis for this subspace is formed by the Dicke states $\ket{D_n}$, $n=0,\ldots,4$, namely the normalised equal superpositions of all computational basis states of the four copies with $n$ qubits in $\ket{1}$, in which $H_4=\frac{1}{5}\sum_n\ket{D_n}\bra{D_n}$.

The six poles are of two kinds.
The poles along $z$ give $\ket{0}^{\otimes4}=\ket{D_0}$ and $\ket{1}^{\otimes4}=\ket{D_4}$.
The other four poles are $(\ket{0}+\theta\ket{1})/\sqrt{2}$ with $\theta\in\{1,-1,i,-i\}$, for which
\eq{
\left[\frac{\ket{0}+\theta\ket{1}}{\sqrt{2}}\right]^{\otimes4}=\frac{1}{4}\sum_{n=0}^4\theta^n\sqrt{\binom{4}{n}}\,\ket{D_n}\,.
\label{eq:sm-dicke}
}
Summing the corresponding projectors over $\theta$ involves $\sum_\theta\theta^n(\theta^*)^{n'}$, which equals $4$ if $n-n'$ is a multiple of $4$ and vanishes otherwise.
The only off-diagonal elements that survive therefore connect $\ket{D_0}$ and $\ket{D_4}$, and including the poles along $z$,
\eq{
S_4=&\frac{1}{4}\ket{D_2}\bra{D_2}+\frac{1}{6}\left(\ket{D_1}\bra{D_1}+\ket{D_3}\bra{D_3}\right)\nonumber\\
&+\frac{5}{24}\left(\ket{D_0}\bra{D_0}+\ket{D_4}\bra{D_4}\right)\nonumber\\
&+\frac{1}{24}\left(\ket{D_0}\bra{D_4}+{\rm h.c.}\right)\,.
\label{eq:sm-S4}
}
The last two terms are diagonalised by $\ket{D_\pm}=(\ket{D_0}\pm\ket{D_4})/\sqrt{2}$, with eigenvalues $5/24\pm1/24$.
$S_4$ therefore has only two distinct eigenvalues in the symmetric subspace,
\eq{
S_4=\frac{1}{4}P_E+\frac{1}{6}P_T\,,
\label{eq:sm-S4-eig}
}
where $P_E$ projects onto the two states $\ket{D_2}$ and $\ket{D_+}$, and $P_T$ onto the three states $\ket{D_1}$, $\ket{D_3}$ and $\ket{D_-}$, which is consistent with ${\rm Tr}\,S_4=1$.
This structure is dictated by symmetry, as single-qubit Cliffords permute the poles and the symmetric subspace carries spin $2$, which splits into the irreducible representations $E$ and $T_2$ of the octahedral group, as do the $d$ orbitals in a cubic crystal field~\cite{tinkham2003group}.

We can now read off the distances.
With $P_\perp$ the projector onto the eleven-dimensional complement of the symmetric subspace, $H_4=(P_E+P_T)/5$ and $I_4=(P_\perp+P_E+P_T)/16$.
Using Eq.~\eqref{eq:sm-S4-eig} in Eq.~\eqref{eq:sm-te-avg}, $\braket{\rho^{(4)}_{\rm TE}}=(q/16)P_\perp+[q/16+(1-q)/4]P_E+[q/16+(1-q)/6]P_T$, such that
\eq{
\braket{\rho^{(4)}_{\rm TE}}-H_4&=\frac{q}{16}P_\perp+\frac{4-15q}{80}P_E-\frac{8+25q}{240}P_T\,,\nonumber\\
\braket{\rho^{(4)}_{\rm TE}}-I_4&=(1-q)\left[\frac{3}{16}P_E+\frac{5}{48}P_T-\frac{1}{16}P_\perp\right]\,.
\label{eq:sm-diff4}
}
As for $k\leq3$, half the trace norm is the sum of the positive eigenvalues~\cite{nielsen-chaung-book}.
The projectors $P_\perp$, $P_E$ and $P_T$ have ranks $11$, $2$ and $3$.
In the first line of Eq.~\eqref{eq:sm-diff4}, $P_\perp$ always contributes $11q/16$ and $P_E$ contributes $(4-15q)/40$ only for $q<4/15$, whereas in the second line, $P_E$ and $P_T$ contribute $(1-q)(3/8+5/16)$.
This leads to
\eq{
{\cal D}^{(4)}_{\rm Haar}&=\max\left(\frac{1}{10}+\frac{5q}{16},\frac{11q}{16}\right)\,,\nonumber\\
{\cal D}^{(4)}_{\mathbb{I}}&=\frac{11(1-q)}{16}\,,
\label{eq:sm-D4}
}
which is Eq.~\eqref{eq:Dk-clifford-4}.
In the pure phase, $q=0$, the TE thus remains at a distance $\frac{1}{2}\norm{S_4-H_4}_1=1/10$ from the Haar ensemble, which is the saturation value in Fig.~\ref{fig:cliffordTE}(c,d).
Finally, tracing out $k-4$ copies maps $S_k\mapsto S_4$ and $H_k\mapsto H_4$, and the trace distance does not increase under partial traces~\cite{nielsen-chaung-book}, such that $\frac{1}{2}\norm{S_k-H_k}_1\geq1/10$ for all $k\geq4$.
The TE is therefore not a $k$-design for any $k\geq4$.

The gap is a direct consequence of the absence of magic.
For a pure state of a qubit, the stabiliser 2-R\'enyi entropy $M_2$ is defined via $2^{-M_2}=\frac{1}{2}(1+\sum_ar_a^4)$~\cite{leone2022stabilizer}, such that the operator $\sum_a(\sigma^a)^{\otimes4}$ measures it on four copies,
\eq{
{\rm Tr}\left[\rho({\bf r})^{\otimes4}\sum_a(\sigma^a)^{\otimes4}\right]=\sum_ar_a^4=2^{1-M_2}-1\,.
\label{eq:sm-magic}
}
This equals $1$ for all stabiliser states, for which $M_2=0$, and $3/5$ on average over the Haar ensemble.
No ensemble of pure stabiliser states can therefore be a 4-design~\cite{zhu2016clifford}.

\section{Ensemble transitions across the forced-measurement transition in all-to-all circuits\label{sec:sm-alltoall}}

In the main text, we had presented results for 1+1D brickwork circuits.
We complement those results in this section by presenting analogous results for the TE and the PE for all-to-all circuits with forced measurements~\cite{nahum2021measurement}.

In these circuits, at each step a two-qubit gate is applied on a randomly chosen pair of qubits with probability $1-r$ and a randomly chosen qubit is subjected to forced measurement (projected up in the computational basis irrespective of the Born-rule probability) with probability $r$. One time step is composed of $N$ such operations.
Note that this naturally means that each trajectory with a certain number of measurements has exactly the same probability.

The forced-measurement purification transition (FMPT) is again a transition from a mixed phase for $r<r_c$, in which $R$ remains entangled with $S$, to a pure phase for $r>r_c$, in which the measurements disentangle and purify $R$, with $r_c=(212+75\pi)/(362+75\pi)\approx0.749$.
The entanglement vanishes on approaching $r_c$ from below with a BKT-like essential singularity, controlled by a correlation length $\xi$ which diverges as $\xi=a\,e^{b/\sqrt{r_c-r}}$~\cite{nahum2021measurement}.
We find that the TE and the PE in this case also show the same ensemble transitions as for the brickwork circuits except they inherit the BKT-like scaling of the FMPTs.

\begin{figure}[t]
    \centering
    \includegraphics[width=\linewidth]{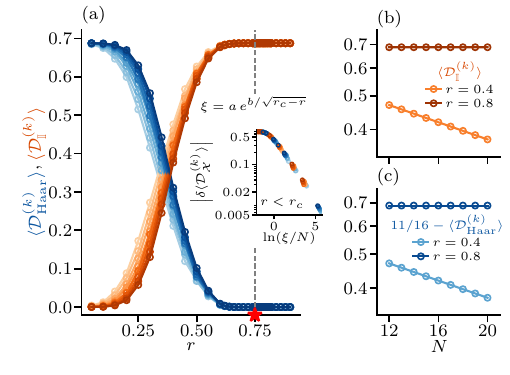}
    \caption{Ensemble transition in the TE across the FMPT in all-to-all circuits with Haar-random gates.
    (a) The distances $\braket{{\cal D}^{(k)}_{\rm Haar}}$ (blue tones) and $\braket{{\cal D}^{(k)}_{\mathbb{I}}}$ (orange tones) of the TE from the Haar and the identity ensembles, Eq.~\eqref{eq:Dk-te}, as a function of $r$ for $k=4$; darker shades denote larger $N$, with $N=12,13,\ldots,20$.
    The dashed line (star) marks $r_c\approx0.749$~\cite{nahum2021measurement}.
    Inset: $|\delta\braket{{\cal D}^{(k)}_{\cal X}}|$, Eq.~\eqref{eq:sm-delta}, with ${\cal X}\in\{{\rm Haar},\mathbb{I}\}$ denoting the Haar and the identity ensembles, for $r<r_c$ as a function of $\ln(\xi/N)$, for $\xi = a\exp[b/\sqrt{r_c-r}]$ with $a\approx0.004$ and $b\approx5.3$.
    (b) $\braket{{\cal D}^{(k)}_{\mathbb{I}}}$ as a function of $N$ for $r=0.4$ ($r<r_c$), where it decays with $N$, and for $r=0.8$ ($r>r_c$), where it remains pinned at $11/16$.
    (c) Same as (b) for $11/16-\braket{{\cal D}^{(k)}_{\rm Haar}}$, as for $r<r_c$ the TE approaches the identity ensemble, which is itself at a distance $11/16$ from the Haar ensemble.}
    \label{fig:Dkvsr-alltoall}
\end{figure}

\subsection{Trajectory ensemble} We plot the distances of the TE from the two ensembles it can approach, $\braket{{\cal D}^{(k)}_{\cal X}}$ with ${\cal X}\in\{{\rm Haar},\mathbb{I}\}$, for $k=4$ [Fig.~\ref{fig:Dkvsr-alltoall}(a)].
In the mixed phase, $R$ remains maximally mixed for almost all trajectories and the TE approaches the identity ensemble, such that $\braket{{\cal D}^{(k)}_{\mathbb{I}}}\to0$ and $\braket{{\cal D}^{(k)}_{\rm Haar}}\to1-(k+1)/2^k=11/16$.
In the pure phase, $R$ is purified and the forced outcomes distribute the resulting pure states such that the TE approaches the Haar ensemble, and the roles of the two distances are reversed.
As $\xi$ is defined only for $r<r_c$, we plot the deviation of the distances from values they saturate to in the pure phase,
\eq{
\delta \braket{{\cal D}^{(k)}_{\cal X}}=\braket{{\cal D}^{(k)}_{\cal X}}^{\rm sat}-\braket{{\cal D}^{(k)}_{\cal X}}\,,
\label{eq:sm-delta}
}
with $\braket{{\cal D}^{(k)}_{\rm Haar}}^{\rm sat}=0$ and $\braket{{\cal D}^{(k)}_{\mathbb{I}}}^{\rm sat}=11/16$.
The data collapses as a function of $\ln(\xi/N)$ (inset of Fig.~\ref{fig:Dkvsr-alltoall}(a)), such that the ensemble transition of the TE coincides with the FMPT and inherits its BKT-like essential singularity.

Fig.~\ref{fig:Dkvsr-alltoall}(b,c) shows how the TE approaches these ensembles with $N$.
For $r>r_c$, it is already Haar for all $N$, and both distances are pinned at their saturation values.
For $r<r_c$, the TE approaches the identity ensemble, such that $\braket{{\cal D}^{(k)}_{\mathbb{I}}}$ decays to zero with $N$ (Fig.~\ref{fig:Dkvsr-alltoall}(b)).
$\braket{{\cal D}^{(k)}_{\rm Haar}}$ instead approaches $11/16$, the distance between the identity and the Haar ensembles, and the quantity that decays is therefore $11/16-\braket{{\cal D}^{(k)}_{\rm Haar}}$ (Fig.~\ref{fig:Dkvsr-alltoall}(c)).

\begin{figure}[t]
    \centering
    \includegraphics[width=\linewidth]{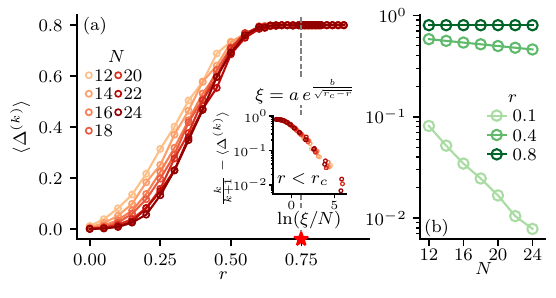}
    \caption{Deep thermalisation across the FMPT in all-to-all circuits with Haar-random gates.
    (a) The trajectory-averaged distance $\braket{\Delta^{(k)}}$ of the PE on $R$ from the Haar ensemble as a function of $r$ for $k=4$ and different $N$.
    The dashed line (star) marks $r_c\approx0.749$~\cite{nahum2021measurement}, beyond which the data saturates to $k/(k+1)=4/5$, the value for the trivial ensemble.
    Inset: Approach to the saturation value, $k/(k+1)-\braket{\Delta^{(k)}}$, for $r<r_c$ as a function of $\ln(\xi/N)$, for $\xi = a\exp[b/\sqrt{r_c-r}]$ with $a\approx0.006$ and $b\approx5.4$.
    (b) $\braket{\Delta^{(k)}}$ as a function of $N$ for $r=0.1$ and $r=0.4$ ($r<r_c$), where it decays with $N$, and for $r=0.8$ ($r>r_c$), where it remains pinned at $4/5$.}
    \label{fig:deltakvspalltoall}
\end{figure}

\subsection{Projected ensemble and deep thermalisation}

We now resolve each member of the TE further by measuring $S$ at the final time, [see Eq.~\eqref{eq:pe-def}], and ask whether the outcomes generate a Haar ensemble on $R$.
The behaviour of the TE above sets the expectations for the two phases.
In the mixed phase, $\rho_R({\bf m})\approx\mathbb{I}/2$ indicating that $R$ is entangled with $S$ and hence the boundary measurements can unravel $\rho_R({\bf m})=\mathbb{I}/2$ into a Haar ensemble with $\Delta^{(k)}({\bf m})\to0$.
In the pure phase, $R$ is purified and disentangled from $S$, so that the boundary outcomes carry no information about $R$ and the PE collapses onto a single state with $\Delta^{(k)}({\bf m})=k/(k+1)$.
We therefore plot $\braket{\Delta^{(k)}}$, Eq.~\eqref{eq:Deltak-m}, which distinguishes the two cases, as a function of $r$ in Fig.~\ref{fig:deltakvspalltoall}(a).

For $r>r_c$, the $\langle\Delta^{(k)}\rangle$ for all $N$ lie on $4/5$, the value for the trivial ensemble when $k=4$.
For $r<r_c$, on the other hand, $\braket{\Delta^{(k)}}$ decays with $N$ [Fig.~\ref{fig:deltakvspalltoall}(b)], such that the PE approaches the Haar ensemble with increasing $N$.
This decay is faster at $r=0.1$, where $\xi\approx5$ is smaller than $N$, than at $r=0.4$, where $\xi\approx60$.
As for the TE, $\xi$ is defined only for $r<r_c$, and in place of the power-law scaling with $(r-r_c)N^{1/\nu}$ of the 1+1D circuits, we plot the approach to the saturation value, $k/(k+1)-\braket{\Delta^{(k)}}$, which collapses as a function of $\ln(\xi/N)$ (inset of Fig.~\ref{fig:deltakvspalltoall}(a)).
The ensemble transitions of the TE and the PE therefore both coincide with the FMPT, as they do with the MIPT in the 1+1D circuits, and inherit its BKT-like essential singularity.

\bibliography{refs}

\end{document}